\documentclass[aps,pra,twocolumn,groupedaddress,fixfloats]{revtex4-1}
\usepackage{graphicx}
\usepackage{amsmath}
\usepackage{amssymb}
\usepackage{bm} 

\begin{document}
\newcommand{\PRA}{\textit{Phys. Rev.} A }
\newcommand{\PRL}{\textit Phys. Rev. Lett.}
\newcommand{\bit}{\bibitem}
\newcommand{\tbf}{\textbf}
\newcommand{\tit}{\textit}
\newcommand{\JPCS}{\textit{J. Phys.} C: Conf. Ser. }
\newcommand{\JPB}{\textit{J. Phys.} B }
\newcommand{\etal}{\textit et. al.}
\title{On Resonance Enhancement of $E1-E2$ Nondipole Photoelectron Asymmetries in Low-Energy Ne $2p$-Photoionization}
\author{Valeriy K. Dolmatov}
\affiliation{Department of Chemistry and Physics, University of North Alabama, Florence, Alabama 35632, USA}
\email[Send e-mail to:\ ]{vkdolmatov@una.edu}
\author{Steven T. Manson}
\affiliation{Department of Physics and Astronomy, Georgia State University, Atlanta, Georgia 30303, USA, smanson@gsu.edu}

\date{\today}

\begin{abstract}
Earlier, a significant enhancement of the nondipole parameters $\gamma_{2p}$, $\delta_{2p}$, and $\zeta_{2p} = \gamma_{2p} + 3\delta_{2p}$ in the photoelectron angular distribution for Ne $2p$ photoionization, owing to resonance interference between dipole ($E1$) and quadrupole ($E2$) transitions, was predicted. This enhancement manifests as narrow resonance spikes in the parameters due to the low-energy $2s \rightarrow 3p$ and $2s \rightarrow 4p$ dipole, as well as the $2s \rightarrow 3d$ quadrupole autoionizing resonances. Given the unique nature of this predicted enhancement, it requires further validation. Specifically, whether these narrow spikes in $\gamma_{2p}$, $\delta_{2p}$, and $\zeta_{2p}$ will or will not retain their values  for experimental observation if one accounts for a typical finite frequency spread in the ionizing radiation.
To address this, we revisit the previous study, now incorporating the effect of frequency spread in the ionizing radiation, assuming a spread as large as $5$ meV at the half-maximum of the radiation's intensity. We demonstrate in the present paper that while the frequency spread does affect the resonance enhancement of $\gamma_{2p}$, $\delta_{2p}$, and $\zeta_{2p}$, these parameters still retain quantitatively significant values to be observed experimentally.
 The corresponding calculations were performed using the random phase approximation with exchange, which accounts for interchannel coupling in both dipole and quadrupole photoionization amplitudes.
\end{abstract}

\maketitle

\section{Introduction}

Understanding both the qualitative and quantitative aspects of the interference between electric dipole ($E1$) and electric quadrupole ($E2$) transitions in the photoionization process is important not only for the interpretation of angle-resolved photoelectron spectroscopy but also from a fundamental point of view, as it offers a unique tool for developing, testing, and refining many-body theories of atomic physics that account for nondipole excitation channels in low-photon-energy ionization processes.
The $E1$-$E2$ interference arises from  the first-order
correction term $i{\bf k\cdot r}$ to the dipole approximation for the
photoionization matrix element between initial and final
states:
$M_{if} = \langle f | (1 + i \bf{k} \cdot \bf{r}) \bf{e} \cdot \hat{p} | i \rangle$,
where ${\bf k}$ and ${\bf e}$ are the
photon momentum  and polarization vector, and ${\bf r}$
and $\hat{\bf p}$ are the electron position vector and momentum operator,
respectively \cite{Amusia1975,Pratt,Cooper93,Shaw} (and references therein). Such studies have particularly gained momentum after the breakthrough in experimental techniques
that have made these interference effects well observable \cite{KraessigPRL95,KraessigPRA96,Hemmers97,Krause} (and references therein). Their unexpectedly strong significance, both predicted and measured, in angle-resolved spectra of photoelectrons
at photon energies not only few keV but also few tens of eV  has led to the discovery of the breakdown of the dipole approximation,
which was considered the only significant ``player'' at such energies. Even nondipole effects from higher-order electric-octupole  and pure-electric-quadrupole excitation channels have been predicted and experimentally confirmed to be significant at photon energies ranging from only $100$ eV to $1500$ keV, in some cases \cite{DereviankoPRL2000}.  The reader is referred, e.g., to a comprehensive review paper \cite{HemmersRPC04} as well as a recent topical review \cite{China}, where more details and references are presented. Here, we only note that initially focused on photoelectron angular asymmetries in one-electron single-photon photoionization of atoms, research on $E1$-$E2$ interference has expanded in recent years to cover its impact on a variety of phenomena, including, to name a few, harmonic generation \cite{China}, double-electron photoionization \cite{Manakov}, photoionization time delays \cite{Mandal,Amusia2020}, ionization by twisted radiation \cite{Gryzlova2023}, sequential two-photon atomic double ionization \cite{Gryzlova2012,Gryzlova2013}, and strong-field atomic ionization \cite{StrongField} (and references therein).

Low-energy angle-resolved photoemission, thus, has been under continuing scrutiny by theorists and experimentalists over a period of several decades and continues to be of interest.

In the present paper, we re-evaluate the previously predicted \cite{DolmMansPRL} $2s \rightarrow 3p$ and $2s \rightarrow 4p$ dipole, as well as the $2s \rightarrow 3d$ quadrupole, autoionizing resonances in the nondipole photoelectron angular asymmetry parameters (referred to as the $\gamma_{n\ell}$, $\delta_{n\ell}$, and $\zeta_{n\ell}$ parameters, as defined in \cite{Cooper93}) for Ne $2p$ photoionization. While the corresponding resonance spikes in $\gamma_{2p}$, $\delta_{2p}$, and $\zeta_{2p}$ were predicted to be significant, they were also narrow, and previous calculations did not account for the frequency spread in the ionizing radiation. The question of how such frequency spread might affect these resonance spikes -- e.g., ``could it potentially wipe the spikes out of the spectrum, thus obliterating nondipole effects in the angular asymmetry of $2p$ photoelectrons?'' -- has remained unexplored until now.

It is the aim of the present paper to address this gap in knowledge by performing a long-overdue refined calculation of the nondipole $\gamma_{2p}$, $\delta_{2p}$, and $\zeta_{2p}$ angular asymmetry parameters in Ne $2p$ photoionization through the region of the $2s \rightarrow 3p$, $2s \rightarrow 4p$, and $2s \rightarrow 3d$ resonances, now incorporating the effects of frequency spread in the photon beam. We demonstrate that the frequency spread in the ionizing radiation does quantitatively affect the resonance spikes in $\gamma_{2p}$, $\delta_{2p}$, and $\zeta_{2p}$. Nevertheless, the spikes remain sufficiently strong to be experimentally detected.

We note that there are examples of calculated $\gamma_{n\ell}$, $\delta_{n\ell}$, and $\zeta_{n\ell}$ in the region of dipole and quadrupole resonances where the resonance spikes in these parameters are significantly wider and quantitatively larger than those observed in Ne. Notable examples include the gigantic resonance enhancements of these parameters near the $3p \rightarrow 3d$ dipole giant autoionizing resonance in Cr and Mn, as well as the $4p \rightarrow 4d$ resonance in Mo and Tc \cite{DolmMn}, or, especially, near the $3s \rightarrow 3d$ giant quadrupole resonance in $3p$ photoionization of Ca \cite{DolmCa}. In the latter case, the nondipole parameters were found to increase drastically, reaching as much as $65\%$ of the dipole counterpart. However, these remarkable examples relate to atoms in the metallic group of the periodic table, and as we understand it, experiments with metallic atom vapors are difficult to perform. In contrast, Ne is a noble gas, for which conducting experiments is  easier. This is why we focus on the photoionization of Ne in the present work.

Corresponding calculations were performed in the framework of the \tit{random phase approximation with exchange} (RPAE) \cite{Amusia_book} with accounting for
interchannel coupling in both dipole and quadrupole photoionization
amplitudes.

Atomic units (a.u.) ($|e|=m_{\rm e}=\hbar = 1$), where $e$ and $m_{\rm e}$ are the electron's charge and mass, respectively, are used throughout the paper unless specified otherwise.

\section{Theory}

To calculate the Ne $\sigma_{n\ell}$ photoionization cross section and the dipole  $\beta_{n\ell}$ as well as the nondipole
$\gamma_{n\ell}$ and $\delta_{n\ell}$ angular-asymmetry parameters, we use well known formulas.  In a one-electron approximation, for $100\%$ linearly polarized light, the angle-differential photoionization cross section, $\frac{d\sigma_{n\ell}}{d\Omega}$, is given
by \cite{Cooper93}:

\begin{eqnarray}
\frac{d \sigma_{n\ell}}{d \Omega}=
\frac{\sigma_{n\ell}}{4 \pi} \left[
1 + \frac{\beta_{n\ell}}{2}(3 \cos^{2}\theta -1)\right]+ \Delta E_{12}.
\label{eqsigma}
\end{eqnarray}
Here, $d\Omega$ is a solid angle, $\sigma_{n\ell}$ is the dipole photoionization cross section of the $n\ell$-subshell, and  $\Delta E_{12}$ is the $E1$-$E2$ interference correction term:
\begin{eqnarray}
\Delta E_{12}= \frac{\sigma_{n\ell}}{4\pi}(\delta_{n\ell}+
\gamma_{n\ell}\cos^{2}\theta)\sin\theta \cos\phi.
\label{eqE_{12}}
\end{eqnarray}
Here, the spherical angles $\theta$ and $\phi$ are defined in relation to
directions of the photon momentum {\bf k}, photoelectron momentum {\bf p},
and photon polarization vector {\bf e}.

The $\sigma_{n\ell}$, $\beta_{n\ell}$, $\gamma_{n\ell}$, and $\delta_{n\ell}$ are, in turn,
given by \cite{Cooper93}
\begin{eqnarray}
\sigma_{n\ell} = && \frac{4 \pi^2 \alpha}{3(2\ell+1)} N_{n\ell}\omega [\ell d^{2}_{\ell-1}+ (\ell+1)d^2{\ell+1}],\\
\beta_{n\ell} = &&
\frac{
\ell (\ell-1) d^{2}_{\ell-1}+(\ell+1)(\ell+2) d^{2}_{\ell+1}
}
{
(2\ell+1)[l d^{2}_{\ell-1} + (\ell+1) d^{2}_{\ell+1}]
} \nonumber \\
&& - \frac{ 6\ell (\ell+1)d_{\ell-1}d_{\ell+1}\cos(\xi_{\ell+1}-\xi_{\ell-1}) } {
(2\ell+1)[\ell d^{2}_{\ell-1} + (\ell+1) d^{2}_{\ell+1}]
},\\
\gamma_{n\ell}=&&
\frac{3 k}{2[\ell d^{2}_{\ell-1}+(\ell+1)d^{2}_{\ell+1}]} \nonumber \\
&& \times \sum\limits_{\ell',\ell''}A_{\ell',\ell''}d_{\ell'}q_{\ell''}\cos(\xi_{\ell''}-
\xi_{\ell'}),
\\
\delta_{n\ell}=&&
\frac{3 k}{2[\ell d^{2}_{\ell-1}+(\ell+1)d^{2}_{\ell+1}]} \nonumber \\
&& \times \sum\limits_{\ell',\ell''}B_{\ell',\ell''}d_{\ell'}q_{\ell''}\cos(\xi_{\ell''}-
\xi_{\ell'}).
\label{eqgamma}
\end{eqnarray}
Here, $\alpha$ is the fine structure constant, $N_{n\ell}$ is the number of the electrons initially in the ionized $n\ell$ subshell of the atom, $\omega$ and $k$ are the photon energy and momentum, respectively, and  $d_{\ell'}$ and $q_{\ell''}$ are the radial dipole and quadrupole
photoionization amplitudes, respectively:
\begin{eqnarray}
d_{\ell'}=\int_{0}^{\infty}{P_{\epsilon \ell'}(r) r P_{n\ell}(r) dr, }
\\
q_{\ell''}=\int_{0}^{\infty}{P_{\epsilon \ell''}(r) r^{2} P_{n\ell}(r) dr}.
\label{eqdq}
\end{eqnarray}
Here, $\ell' = \ell \pm 1$, $\ell'' = \ell, \ell\pm 2$, $P_{n\ell}(r)/r$ and $P_{\epsilon \lambda}(r)/r$ are the radial
parts of the electron wave functions in the bound $n\ell$ state and in
the continuous $\epsilon \lambda$ spectrum, respectively, $\xi_{\lambda}$ are the phase shifts of the
wave functions of photoelectrons in the field of the positive ionic
core, and the coefficients $A_{\ell',\ell''}$ and $B_{\ell',\ell''}$ depend on a combinations of only orbital quantum numbers $\ell'$ and $\ell''$ and
are tabulated in \cite{Cooper93}.

In the present work, we account for electron correlation in the form of various initial-state and final-state interchannel
couplings between the dipole $2s \rightarrow \epsilon(n) p$ and
$2p \rightarrow \epsilon(n) d$ and $2p \rightarrow \epsilon(n) s$, on the one hand, and, on the other hand,
between the quadrupole $2s \rightarrow \epsilon(n) d$ and $2p \rightarrow \epsilon(n) f$
and $2p \rightarrow \epsilon(n) p$
transitions into discrete and continuum spectra of Ne. To meet the goal, we employ a non-relativistic
RPAE \cite{Amusia_book}. We choose RPAE, because it is a well established method,
which has been  used in atomic studies with a great success for decades.

In RPAE, the dipole and quadrupole matrix elements become complex. To account for this,
the following substitution must be made in the above equations for $\sigma_{n\ell}$, $\beta_{n\ell}$,
$\gamma_{n\ell}$, and $\delta_{n\ell}$ \cite{Amusia2001}:
\begin{eqnarray}
d_{\lambda}^{2} \rightarrow |D_{\lambda'}|^{2},\nonumber \\
d_{\lambda}w_{\lambda'}\cos\Delta\xi \rightarrow && \nonumber \\
&& (D'_{\lambda'}W'_{\lambda'} + D''_{\lambda'}W''_{\lambda'})
\cos\Delta\xi \nonumber \\
&& +(D''_{\lambda'}W'_{\lambda'} - D'_{l'}W''_{\lambda'})
\sin\Delta\xi, \nonumber \\
\Delta\xi=\xi_{\lambda'}-\xi_{\lambda}.
\label{eqDQ}
\end{eqnarray}
Here, $w$ ($W$) stands either for a quadrupole matrix element $q$ and $Q$,
or a dipole matrix element $d$ and $D$, calculated in a Hartree-Fock (HF) or RPAE approximation, respectively,
$D'$ and $Q'$ are real parts of corresponding matrix elements,
whereas $D''$ and $Q''$ are their imaginary parts.

\begin{figure}
\includegraphics[width=7cm]{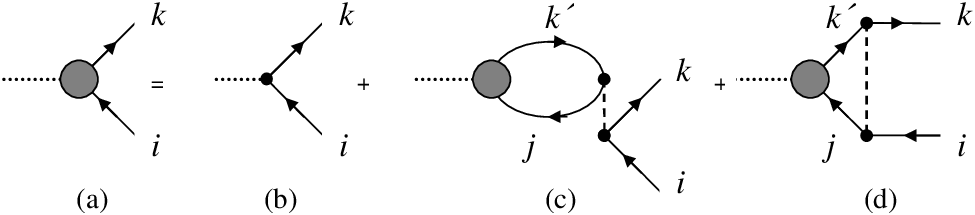}
\caption{Feynman
Diagrammatic representation of the RPAE equation for the
photoionization amplitude $\langle k|\hat{W}|i\rangle $ of the \textit{i}'th subshell
into the \textit{k}'th final state \cite{Amusia_book}. Here, lines with arrows the left (right)
correspond to holes (electrons) in the atom, a
dotted line represents an incoming photon, a dashed line represents the
Coulomb interaction $V(r)$ between the atomic electrons, and a shaded circle
in the vertices of
diagrams (a) (c), and (d) marks
 the effective operator $\hat{W}$ for the photon-atom interaction including electron
correlation in the atom.}
\label{fig1}
\end{figure}

The key points of RPAE are as follows (the reader is referred to \cite{Amusia_book} for details).
RPAE uses the HF basis as the vacuum
state.  With the aid of the Feynman
diagrammatic technique, the RPAE photoionization
amplitude $\langle k |W| i\rangle \equiv W_{ik}$ (whether dipole or quadrupole) of the \textit{i}th subshell of an atom is shown in Figure~\ref{fig1}.

Diagrams (c) and (d) represent RPAE corrections to the
HF photoionization amplitude $\langle k|\hat{w}|i\rangle \equiv w_{ik}$
[diagram (b)]. The corresponding RPAE equation for $(k|\hat{W}|i)$ is \cite{Amusia_book}:
\begin{eqnarray}
\langle k|\hat{W}|i\rangle) = \langle k|\hat{w}|i\rangle +
\sum_{j,k'}\langle j|\hat{W}|k'\rangle\chi_{jk'}\langle k'i|U|jk\rangle.
\label{eqRPAE}
\end{eqnarray}
Here,
\begin{eqnarray}
\chi_{jk'}=\frac{1}{\hbar\omega-\omega_{jk'}+i\zeta} -
\frac{1}{\hbar\omega+\omega_{jk'}-i\zeta},
\label{eqkappa}
\end{eqnarray}
\begin{eqnarray}
\langle k'i|U|jk\rangle = \langle k'i|V|jk\rangle-\langle k'i|V|kj\rangle\delta_{\mu_j\mu_i}.
\label{eqU}
\end{eqnarray}
In the above equations, $\omega_{jk'}$ is the excitation energy of the
$j \rightarrow k'$ transition, $V$ is the Coulomb interaction
operator, $\delta_{\mu_j\mu_i}$ is the Kronecker's delta, $\mu_j$
($\mu_i$) is the electron's $z$-spin-projection in a state $j$
($i$), the summation is performed over all possible
core-intermediate-states \tit{j} and excited-virtual-states $k'$
(including integration over the energy of continuous states), and
$i\zeta$ ($\zeta \rightarrow +0$) indicates the path of integration
around the pole in Equation~(\ref{eqRPAE}).

In these calculations, to maintain the equality of length and velocity gauges in the application of the
RPAE, we use the HF values of the ionization potentials, $I_{n\ell}$, of Ne: $I_{2s} \approx 52.53 eV$ and $I_{2p} \approx 23.14$ eV. Therefore,
$\sigma_{2p}$, $\beta_{2p}$, $\gamma_{2p}$, and $\delta_{2p}$, calculated
near the $2s \rightarrow n\ell$ autoionizing resonances should be shifted by a corresponding amount
to match the resonance positions determined experimentally, if needed.

Finally, to calculate $\beta_{n\ell}$, $\gamma_{n\ell}$, $\delta_{n\ell}$, and $\zeta_{n\ell} = \gamma_{n\ell} + 3\delta_{n\ell}$, taking into account the  frequency spread in the ionizing radiation, we use the Gaussian function, $G(\omega - \omega')$, for a normal probability distribution
with dispersion $\sigma^2$ \cite{Korn}:
\begin{eqnarray}
G(\omega-\omega')= \frac{1}{\sigma \sqrt{2\pi}}\exp\left[-  \frac{(\omega-\omega')^2}{2\sigma^2}\right].
\label{Gauss}
\end{eqnarray}
The full-width at half-maximum for this Gaussian is determined by finding the half-maximum points $\omega_0$. This leads to
\begin{eqnarray}
{\rm FWHM} = 2 \sqrt{2 \ln 2\sigma} \approx 2.3548 \sigma.
\end{eqnarray}
Correspondingly, assuming, in our work, that FWHM = $5$ meV, we find that $\sigma \approx 2.1233$ meV.

To apply the frequency distribution to convolute $\beta_{n\ell}$,  $\gamma_{n\ell}$, $\delta_{n\ell}$, and $\zeta_{n\ell}$,
we first need to convolute the angle-differential photoionization cross section, $\frac{d\sigma_{n\ell}}{d\Omega}$,  determined by Equations~(\ref{eqsigma}) and (\ref{eqE_{12}}):
\begin{eqnarray}
\frac{d\sigma^*_{n\ell}}{d\Omega} = \int G(\omega-\omega') \frac{d\sigma_{n\ell}(\omega')}{d\Omega} d \omega'.
\end{eqnarray}
After simple and obvious mathematical derivations, we get, for example, for convoluted $\gamma_{n\ell}$, to be labeled as $\gamma^*_{n\ell}$:
\begin{eqnarray}
\gamma^*_{n\ell}(\omega) = \frac{1}{\sigma^*_{n\ell}(\omega)}\int \sigma_{n\ell}(\omega')\gamma_{n\ell}(\omega')G(\omega-\omega')d\omega'.
\label{gamma*}
\end{eqnarray}
Here, $\sigma^*_{n\ell}$ is a convoluted photoionization cross section:
\begin{eqnarray}
\sigma^*_{n\ell}= \int \sigma_{n\ell}(\omega')G(\omega-\omega')d \omega'.
\label{sigma*}
\end{eqnarray}

Expressions for convoluted $\beta^*_{n\ell}$, $\delta^*_{n\ell}$ and $\zeta^*_{n\ell}$ are similar to Equation~(\ref{gamma*}) with obvious substitutions: $\gamma_{n\ell} \rightarrow \beta_{n\ell}$, $\delta_{n\ell}$, and   $\zeta_{n\ell}$,
respectively.

\section{Results and Conclusion}

\begin{figure}
\includegraphics[width=7cm]{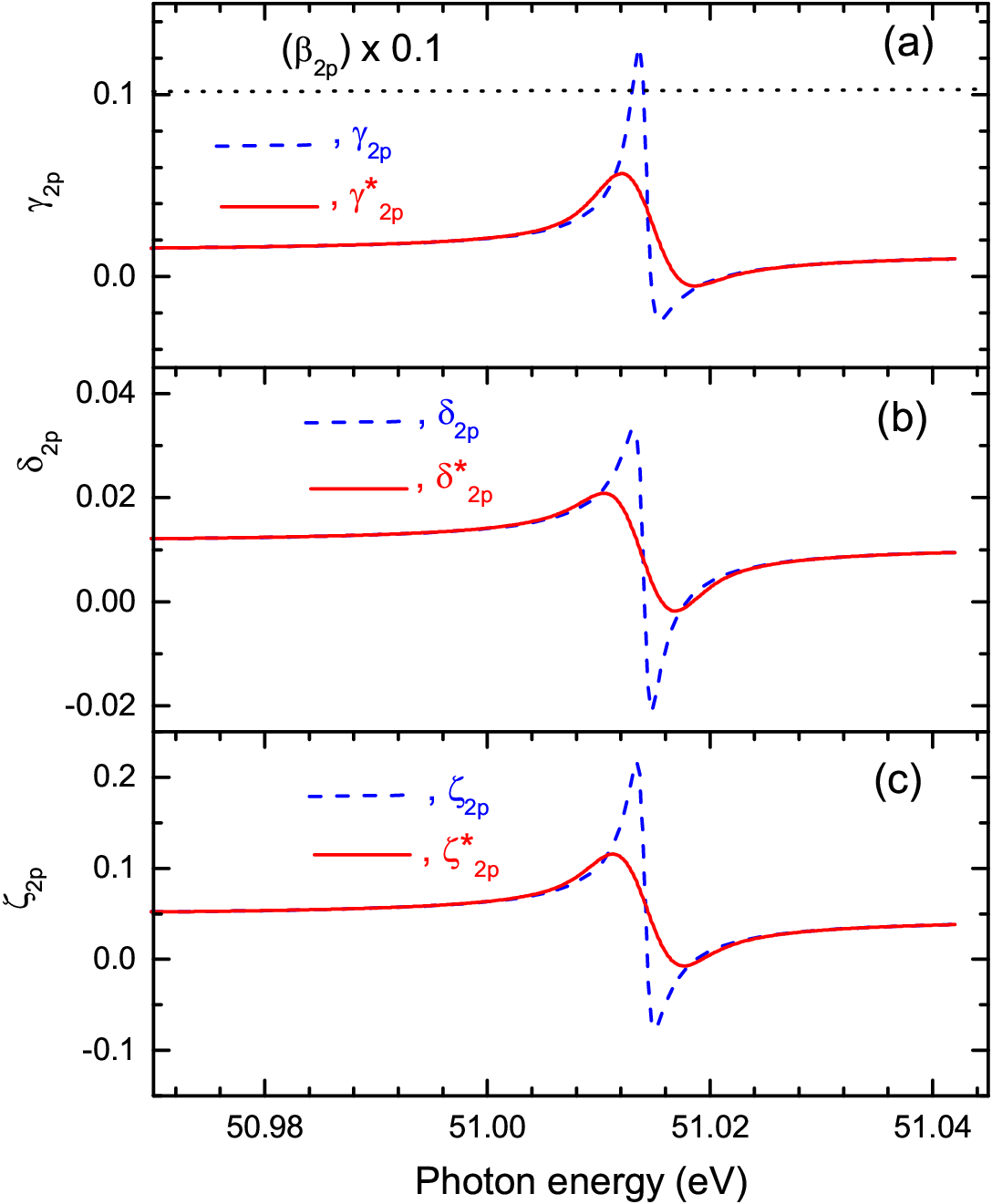}
\caption{RPAE calculated nondipole angular asymmetry parameters for Ne $2p$-photoelectrons through the $2s \rightarrow 3d$ quadrupole autoionizing resonance: (a), $\gamma_{2p}$ and $\gamma^*_{2p}$; (b), $\delta_{2p}$ and $\delta^*_{2p}$; and (c) $\zeta_{2p}$ and $\zeta^*_{2p}$. An asterisk ($^*$) in these plots marks the parameters that were calculated with account  for a $5$ meV FWHM of radiation, as designated in the figure. Also, for comparison) plotted in (a) is the $\beta_{2p}$ dipole angular asymmetry parameter (divided by $10$) which itself is approximately equal to unity.
}
\label{fig2}
\end{figure}

Calculated $\gamma_{2p}$, $\delta_{2p}$, and $\zeta_{2p}$, with and without accounting for FWHM = $5$ meV frequency spread in the ionizing radiation,  are depicted in Figure~\ref{fig2} in the vicinity of the $2s \rightarrow 3d$ quadrupole autoionizing resonance.

One can see that frequency spread has a significant impact on $\gamma_{2p}$, $\delta_{2p}$, and $\zeta_{2p}$. For instance, its account reduces the maximum value of $\gamma_{2p} \approx 0.12$ to $\gamma^*_{2p} \approx 0.06$, which corresponds to approximately $6\%$ of $\beta_{2p}$. Similarly, $\zeta_{2p} \approx 0.22$ decreases to $\zeta^*_{2p} \approx 0.12$, making it about $12\%$ of $\beta_{2p}$ at its maximum.
Despite these reductions, the nondipole angular asymmetries in Ne $2p$-photoelectrons remain within experimental detection limits. Indeed, for example, in \cite{Krause}, even much weaker interference effects -- accounting for only around $0.5\%$ of the dipole contribution -- were successfully observed experimentally.

\begin{figure}
\includegraphics[width=7 cm]{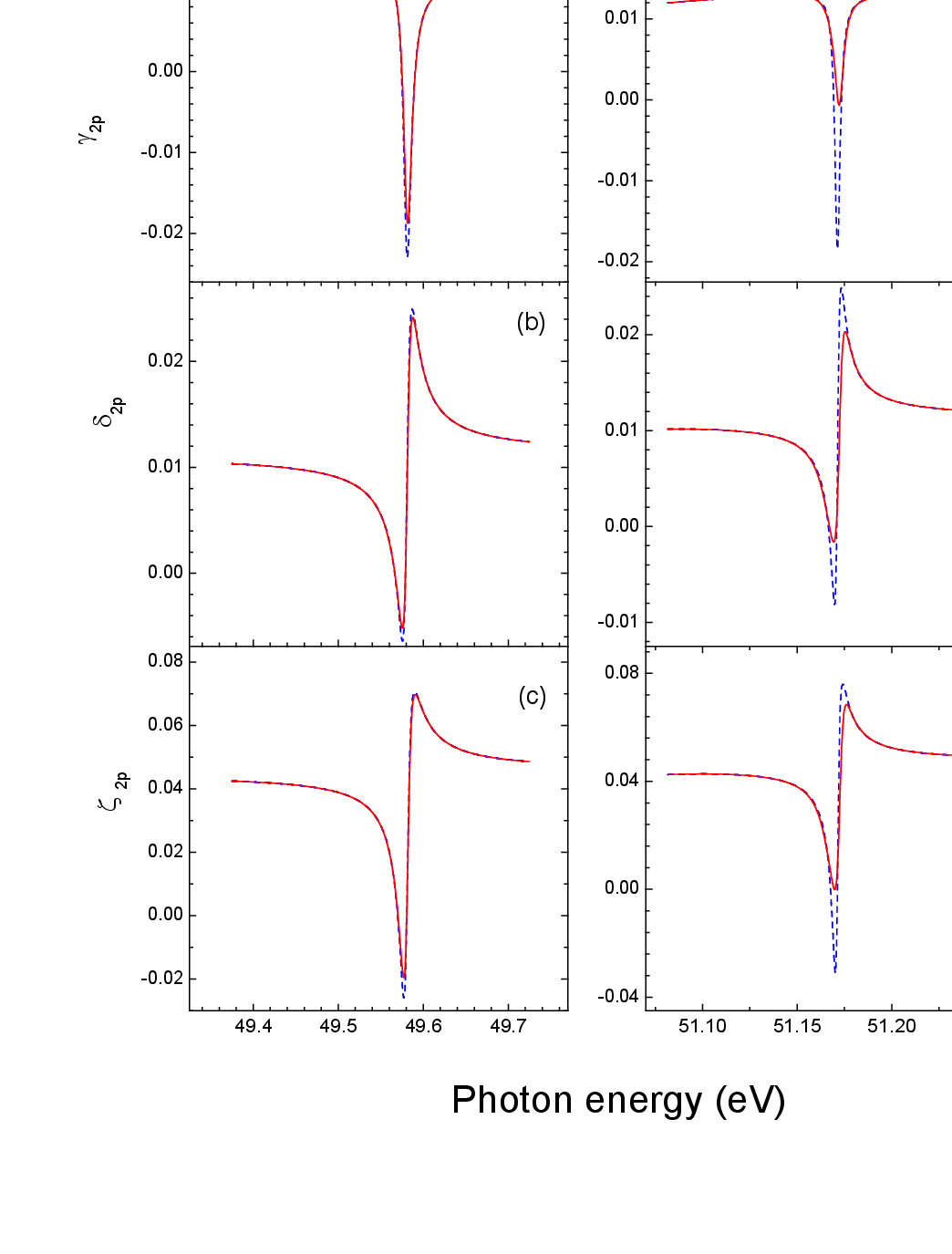}
\caption{RPAE calculated nondipole angular asymmetry parameters for Ne $2p$-photoelectrons through the $2s \rightarrow 3p$ [(a) - (c)] and $2s \rightarrow 4p$ [(d) - (f)] dipole autoionizing resonances. On all plots: solid line shows the parameters calculated with account for the $5$ meV FWHM of radiation,  dash - without such account.}
\label{fig3}
\end{figure}

Figure~\ref{fig3} illustrates the calculated values of $\gamma_{2p}$, $\delta_{2p}$, and $\zeta_{2p}$, with and without accounting for frequency spread in the ionizing radiation, through the region of the $2s \rightarrow 3p$ and $2s \rightarrow 4p$ dipole resonances.

It is evident that the frequency spread in the ionizing radiation has a minimal impact on the $\gamma_{2p}$, $\delta_{2p}$, and $\zeta_{2p}$ parameters in the region of the $2s \rightarrow 3p$ dipole resonance, except at the very tips of the spikes in these parameters. This is because the overall energy widths of the spikes are significantly broader than the $5$ meV frequency spread of the radiation. However, even at the resonance tips, the frequency spread only slightly affects $\gamma_{2p}$, $\delta_{2p}$, and $\zeta_{2p}$.

Next, the effect of frequency spread on $\gamma_{2p}$, $\delta_{2p}$, and $\zeta_{2p}$ becomes more pronounced in the region of the higher-lying, narrower $2s \rightarrow 4p$ resonance, particularly impacting $\gamma_{2p}$. Nevertheless, the resonance enhancement of $\gamma_{2p}$, $\delta_{2p}$, and $\zeta_{2p}$ remains noticeable, even after accounting for the frequency spread.

In conclusion, this paper demonstrates that the $2s \rightarrow 3d$ quadrupole resonance, as well as the $2s \rightarrow 3p$ and $2s \rightarrow 4p$ dipole resonances, in the angular asymmetry parameters $\gamma_{2p}$, $\delta_{2p}$, and $\zeta_{2p}$ for Ne $2p$-photoionization, remain sufficiently large to be experimentally observable, even with a frequency spread of up to $5$ meV in the incident radiation. We  encourage experimentalists to conduct such an experiment.

\begin{acknowledgements}
The work of STM was supported by the US Department of Energy, Office of Basic Sciences, Division of Chemical Science, Geosciences and Biosciences under Grant No. $DE-SC0025316$.

\end{acknowledgements}

\end{document}